\documentclass[preprint2, times]{aastex62}
\usepackage{graphicx}
\usepackage{mathtools, amsmath}
\usepackage{longtable}
\usepackage{url}
\usepackage{color}
\usepackage[utf8]{inputenc}

\shorttitle{Mass-Loss in M-Giants}
\shortauthors{Rau et al.}

\begin{document}


\title{GHRS Observations of Cool, Low-Gravity Stars. \textsc{vi} \\  Mass-Loss Rates and Wind Parameters for M Giants}


\author{Gioia Rau}
\affil{NASA/GSFC Code 667, Goddard Space Flight Center, Greenbelt, MD 20071}
\email{gioia.rau@nasa.gov}

\author{Krister E.~Nielsen}
\affil{Catholic University of America, Washington, DC 20064}

\author{Kenneth G. Carpenter}
\affil{NASA/GSFC Code 667, Goddard Space Flight Center, Greenbelt, MD 20071}

\author{Vladimir Airapetian}
\affil{NASA/GSFC Code 667, Goddard Space Flight Center, Greenbelt, MD 20071}

\begin{abstract}
The photon-scattering winds of M-giants absorb parts of the chromospheric emission lines and produce self-reversed spectral features in high resolution {\it HST}/GHRS spectra. These spectra provide an opportunity to assess fundamental parameters of the wind, including flow and turbulent velocities, the optical depth of the wind above the region of photon creation, and the star's mass-loss rate. This paper is the last paper in the series ``GHRS Observations of Cool, Low-Gravity Stars''; the last several have  compared empirical measurements of spectral emission lines with models of the winds and mass-loss of K-giant and supergiants. We have used the Sobolev with Exact Integration (SEI) radiative transfer code, along with simple models of the outer atmosphere and wind, to determine and compare the wind characteristics of the two M-giant stars, $\gamma$~Cru (M3.5III) and $\mu$~Gem (M3IIIab), with previously derived values for low-gravity K-stars. The analysis specifies the wind parameters and calculates line profiles for the \ion{Mg}{2} resonance lines, in addition to a range of unblended \ion{Fe}{2} lines. Our line sample covers a large range of wind opacities and, therefore, probes a range of heights in the atmosphere. 

Our results show that $\mu$~Gem has a slower and more turbulent wind then $\gamma$~Cru. Also, $\mu$~Gem has weaker chromosphere, in terms of surface flux, with respect to $\gamma$~Cru. This suggests that $\mu$~Gem is more evolved than $\gamma$~Cru. Comparing the two M-giants in this work with previously studied K-giant and supergiant stars ($\alpha$~Tau, $\gamma$~Dra, $\lambda$~Vel) reveals that the M-giants have slower winds than the earlier giants, but exhibit higher mass-loss rates. Our results are interpreted in the context of the winds being driven by Alfv\'{e}n waves.
\end{abstract}


\section{Introduction}

The mass-loss from evolved stars such as red giant, red supergiant, and asymptotic giant branch (AGB) stars, contributes substantially to the chemical enrichment of the Universe \citep{habing04}.  In the cooler, more evolved stars such as AGB stars, the mass-loss is driven by an interplay of pulsation which extends the atmosphere, dust formation, and radiation pressure acting upon the dust grains that eventually leads  to a stellar wind (for a comprehensive discussion on the topic see e.g. \citealp{hofner03} for the theoretical modeling or \citealp{lopez93, danchi94, rau15, rau16, rau17, wittkowski18} for the the comparison to e.g.~interferometric data for objects of various chemistry). In comparison, the chromosphere plays a more critical role in the mass-loss process in warmer stars, such as red giants and supergiants \citep{linsky17}.

Chromospheres and winds are the signature of all cool stars with spectral type later than F5. Evolved stars, unlike main-sequence stars, have chromospheres that are more bloated, with slower and more massive stellar winds. Chromospheres are affected by the mechanical energy flux imparted from the photosphere, and the winds of giant stars are driven by acoustic and magnetic waves generating in those environments \citep{airapetian15, charbonneau95, verdini07}. Earlier studies \citep[e.g.][]{linsky78} on the radiative cooling from emission lines, including \ion{H}{1}~Ly$\alpha$, \ion{Mg}{2}, and \ion{Ca}{2} lines enhanced our knowledge of the mechanism that drives the mass-loss of giant stars, and enabled further studies of  the physical phenomena in the chromospheres. This study includes measurements of the surface fluxes of \ion{Mg}{2} and \ion{Fe}{2} lines (see Section~\ref{resultsSEI}).

This paper is the last one of a series ``GHRS Observations of Cool, Low-Gravity Stars''. Two of the previous papers \citep[e.g][]{robinson98a, carpenter99a} compared  empirical measurements of high-resolution UV spectral emission lines recorded with {\it HST}/GHRS with theoretical models to examine the winds and mass-loss of K-giant and supergiant stars ($\alpha$~Tau, K5III; $\gamma$~Dra, hybrid K5III; and $\lambda$~Vel, K4Ib). These papers presented both empirical measurements of the chromospheric and wind lines in the spectra of cool, evolved stars, along with some exploratory SEI modeling to obtain initial estimates of the wind parameters and mass-loss rates.

The present work completes the series of {\it HST}/ GHRS studies.  We extend the analysis to include two M-giant stars: $\gamma$~Cru (M3.5III) and the slightly more luminous $\mu$~Gem (M3IIIab). This allows us to study the dependence of the wind and mass-loss on spectral type and surface temperature by comparison with the previously studied stars, in addition to surface gravity and luminosity by mutual comparison of the objects analyzed in this paper. Gamma~Cru and $\mu$~Gem are substantially cooler than the K5 stars in the previous studies, but both have sufficiently high effective temperature and low luminosity to allow us to use the Sobolev with Exact Integration (SEI, \citealp{lam87}) methodology. This paper aims to: finalize the analysis of the {\it HST}/GHRS data; provide an initial analysis of the last remaining dataset from that program, utilizing similar techniques for a fair comparison with the preceding studies; present a summary for all of the objects in the series. We would like to stress that sophisticated magnetohydrodynamic (MHD) modeling is the next logical step to enhance our knowledge of these objects, but this purpose is beyond the objectives of this paper.

\begin{figure*}[h]
\includegraphics[width=\hsize, bb=69 79 704 550, angle=180]{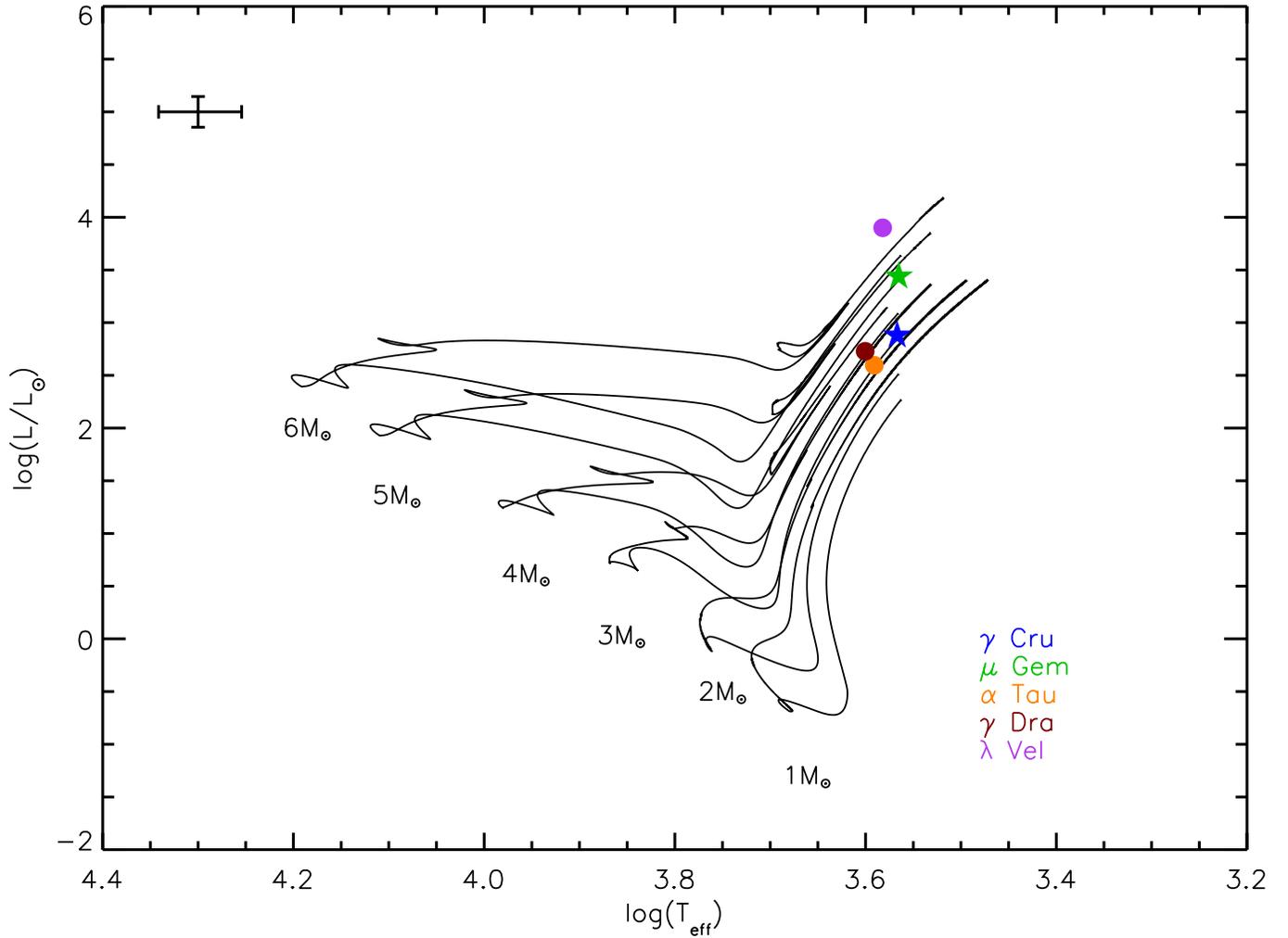}
\caption{Hertzsprung-Russell diagram with evolutionary tracks from \cite{marigo13} for various stellar masses. Different colors indicate the location in the diagram of the various stars. $\gamma$~Cru (\textcolor{blue}{\Large$\star$}) and $\mu$~Gem (\textcolor{green}{\Large$\star$}), while the filled circles represent the comparison stars for the earlier papers in this program. A typical error bar is shown in the upper left side of the figure.}\label{hr-diagram}
\end{figure*}

Mu~Gem is a long-period variable star of M3IIIab spectral type. It has a $V$-band magnitude of 2.87 \citep{ducati02}, at a distance of 71.02$\pm$0.05~pc \citep{vanleeuwen07}. Gamma~Cru is a M3.5III type star, with a $V$-band magnitude of $V$=1.64 \citep{ducati02} at 27.15$\pm$0.55~pc \citep{vanleeuwen07}. The location of the two analyzed M-stars in the Hertzsprung Russell diagram is shown in Figure~\ref{hr-diagram}, which includes evolutionary tracks from \cite{marigo13} for various stellar masses, as well as the K-stars previously studied in this program. The error in luminosity in Figure~\ref{hr-diagram} is assumed to be $\sim$40\% based on the distance uncertainty, while the temperature errors are estimated through the standard propagation of error.

For each star the observations taken with the {\it HST}\ and the data reduction techniques are shown in Section~\ref{obs}. Section~\ref{modeling} describes the modeling techniques, and Section~\ref{results} presents the results and the stellar parameters derived from our modeling. These results are discussed and used to provide constraints on theoretical models of wind acceleration and mass-loss in Section~\ref{discussion}; lastly, Section~\ref{conclusion} presents our conclusions.

\newpage
\section{Observations and Data Reduction \label{obs}}
We have analyzed UV spectra of two late-type M-giant stars, $\gamma$~Cru and $\mu$~Gem, obtained with the GHRS onboard the {\it HST}\ in GTO Programs 1195 and 4685. The observed spectra are summarized in Table~\ref{GHRS-obs}. They cover selected regions in the 2300$-$2850~\AA\ wavelength range that contain a wide variety of chromospheric emission lines of, e.g., \ion{Mg}{2} and \ion{Fe}{2}, which show overlying wind absorption features.

\begin{deluxetable*}{cccccc}
\tablecolumns{6}
\tabletypesize{\scriptsize}
\tablecaption{GHRS Observations of Program Stars. \label{GHRS-obs}}
\tablewidth{0pc}
\tablehead{
\colhead{Obs. Num.} & \colhead{Grating/Slit} & \colhead{Start Time (UT)} & \colhead{ $\Delta\lambda$ (\AA)} & \colhead{Exp.Time (min.)} & \colhead{Disp. (\AA/Diode)}}
\startdata
\multicolumn{6}{l}{$\gamma$~Cru:  March 24, 1992} \\ \hline
 Z0WI0113T & ECH-B/SSA & 23:31:53.56 & 2589.8$-$2602.4 &   29.6 & 0.026 \\
 Z0WI0119T & ECH-B/SSA & 00:51:58.56 & 2791.5$-$2806.5 &  $\sim$6.3 & 0.029 \\
 Z0WI010NT & G270M/SSA & 20:20:37.56 & 2321.1$-$2368.9 &   14.8 & 0.095 \\
 Z0WI010QT & G270M/SSA & 20:42:10.56 & 2476.2$-$2523.4 &   10.0 & 0.093 \\
 Z0WI010TT & G270M/SSA & 21:38:22.56 & 2585.7$-$2632.5 &  $\sim$8.0 & 0.092 \\
 Z0WI010WT & G270M/SSA & 21:53:29.56 & 2731.0$-$2777.1 &  $\sim$4.5 & 0.091 \\
 Z0WI010ZT & G270M/SSA & 22:04:52.56 & 2780.8$-$2826.6 &  $\sim$4.5 & 0.091 \\ \hline
 \multicolumn{6}{l}{$\mu$~Gem:  September 27$-$28, 1993} \\ \hline
 Z1KZ0507T & G270M/SSA &  22:08:33.64 & 2788.0$-$2833.8 &  9.9 & 0.091 \\
 Z1KZ0509T & G270M/SSA &  23:11:27.39 & 2311.4$-$2359.3 & 29.6 & 0.095 \\
 Z1KZ050BN & G270M/SSA &  23:44:58.64 & 2589.4$-$2636.1 & 14.8 & 0.092 \\
 Z1KZ050DT & G270M/SSA &   0:54:46.64 & 2734.7$-$2780.8 & 14.8 & 0.091 \\
 Z1KZ050FT & G270M/SSA &   1:12:57.64 & 2828.4$-$2874.0 & 14.8 & 0.091 
\enddata
\end{deluxetable*}

We have in hand, for comparison, GHRS spectra and analytic results from our previous studies of the warmer non-coronal giant $\alpha$~Tau (K5III), the hybrid K-giant $\gamma$~Dra (K5III), and the K4Ib supergiant $\lambda$~Vel. The observations of the two K-giant stars are described in \cite{robinson98a} and the K-supergiant in \cite{carpenter99a}.

For the two targets of the present study, we use a similar approach. To summarize: we used the Small Science Aperture (SSA) for these observations to optimize the fidelity of the UV line profiles and the precision of the measured radial velocities. Dedicated wavelength calibration exposures of  the on-board platinum lamps (WAVECALS) were obtained close in time to each of the stellar observations to  allow us to determine the dispersion coefficients and absolute wavelengths of each of the stellar observations to an accuracy of better than 0.3 diode widths, corresponding to about $3$~km~s$^{-1}$ in the medium resolution ($R$=$\lambda/\Delta \lambda$=25,000) observations and about 0.5~km~s$^{-1}$\ in the high resolution ($R$=$\lambda/\Delta \lambda$=80,000) observations. We divided long exposures into a series of sub-exposures, each with an integration time of 10 minutes or less to further reduce the effects of thermal drifts and geomagnetic interactions within the
spectrograph (see \citealp{sod93}). These exposures were used during the data reduction process to measure and correct for any such drifts within the spectrograph.

The CALHRS routine developed by the GHRS Investigation Definition Team (IDT) was used to reduce and calibrate the observations.  This program merges the individual samples into a single spectrum, subtracts background counts and corrects for non-linearities in detector sensitivity. It then corrects for vignetting and the echelle blaze function and applies an absolute flux calibration \citep{sod93}. The program used the WAVECAL exposures associated with each science observation, and obtained at the same grating carrousel position, to produce the optimal wavelength calibration. The separate sub-exposures at a given wavelength were then cross-correlated and co-added to produce the final spectra.  We compared these calibrated spectra to the most recent calibrated spectra in the MAST archive and did not find any differences significant enough to impact the measurements and conclusions reported in this paper.

\section{Modeling UV emission lines as wind diagnostics}\label{modeling}
This paper utilize two techniques to derive wind parameters, both using \ion{Mg}{2} $\lambda$$\lambda$2796, 2803 and a large number of \ion{Fe}{2} transitions formed at different height in the stellar atmosphere. First, we use a parametrized fitting procedure of the lines to obtain flow velocities for the emission and the superimposed absorption components. This technique is explained in Section~\ref{empiricalmeasures} and the results are discussed in Section~\ref{results_empirical}. Second, we use a SEI model \citep{lam87} to derive wind parameters such as $v_\infty$, $v_\text{turb}$, and mass-loss rates. The SEI model is explained in Section~\ref{sei_form} and its results in Section~\ref{resultsSEI}.

\subsection{Empirical measures of wind parameters}\label{empiricalmeasures}
Photon-scattering winds of cool, low-gravity stars produce absorption features in the strong chromospheric emission lines. The strength and shape of these wind absorptions are sensitive to the wind opacity, turbulence, and flow velocity. Complementing this, the wings of the emission features, which are not affected by wind absorption, can be used to measure the velocity of the line photon creation region.   

Self-absorptions extend further to the blue of line center in lines of higher opacity, since the last photon scatterings in the more opaque lines occur at higher altitudes, where the accelerating wind is flowing faster than in the weaker lines. Thus, by examining a set of \ion{Fe}{2} and \ion{Mg}{2} self-reversed lines of different strength, with a a set of transitions representing a range of optical depths, we can map the acceleration of the stellar wind.  

\begin{figure*}[h]
\begin{centering}
\includegraphics[width=\hsize, bb=23 8 989 769, angle=0]{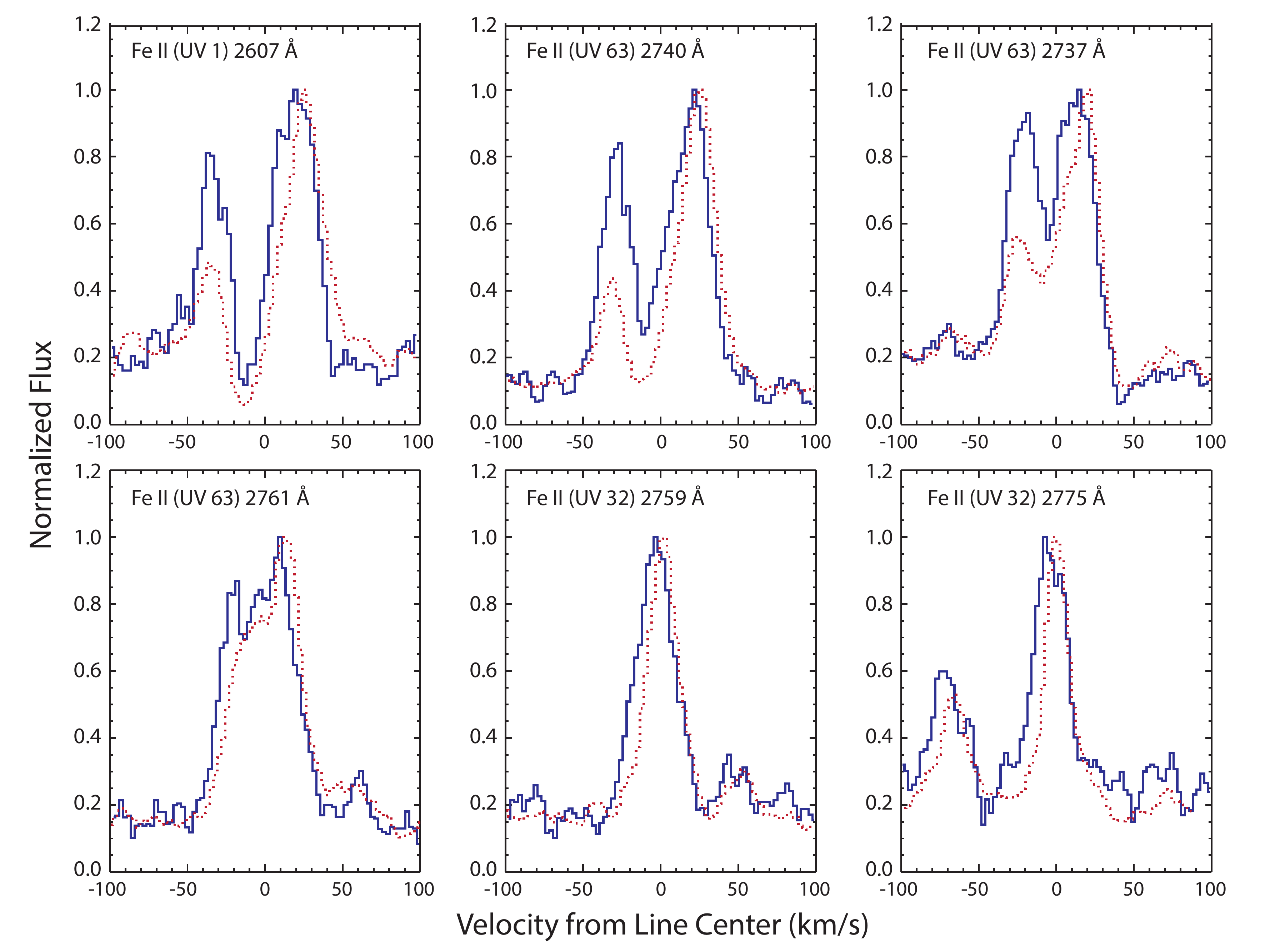}
\caption{Examples of self-reversed Fe~II emission lines in the spectrum of $\mu$~Gem (blue solid line) and $\gamma$~Cru (red dashed line). The spectra are corrected for the objects' radial velocity: $v_r$=54.46 km~s$^{-1}$\  for $\mu$~Gem \citep{massarotti08}, and $v_r$= 20.6~km~s$^{-1}$\ for $\gamma$~Cru \citep{wielen99}. \label{feiimgii}}
\end{centering}
\end{figure*}

The shifting wavelengths of the wind absorptions relative to the emission peaks, and the changes in relative strengths of the emission peaks, reflect the acceleration of the wind above the chromosphere. Figure~\ref{feiimgii} shows examples of such  \ion{Fe}{2}  lines in the spectra of the two M giants. 

The emission line measurements for $\mu$~Gem and $\gamma$~Cru are presented in Table~\ref{FeII_MuGem_GamCru}, including the calculated relative optical depth which is proportional to the line center optical depth. 
Following \cite{judge86b} and~\cite{carpenter95, carpenter99a}, the optical depth $\tau_{\text{rel}}$ is calculated at $T$=6000~K. In this way we can compute a relative optical depth scale, which allows us to order lines according to the actual optical depth of the \ion{Fe}{2} lines in the wind \citep[see][]{carpenter99a}. The Gaussian fits are used, rather than integrating the line profile, to characterize the chromospheric flux, and of the amount of wind absorption. The wind is assumed to be a pure scattering medium to estimate the emission and absorption fluxes. The properties of the unreversed emission lines were determined by fitting a single Gaussian to the observed line profile. 

The parameters of the fit provide estimates of the integrated flux, width, and radial velocity of the line; for the self-reversed lines we adopted an empirical model to parameterize the line. In this model the lines wings are represented by a Gaussian profile which is validated by the quality of the fits. The transition core reversal is assumed to be formed in an overlying reversing layer.

\begin{deluxetable*}{ccc|cccc|cccc}
\tablecolumns{9}
\tablewidth{0pc}
\tablecaption{\label{FeII_MuGem_GamCru} \ion{Fe}{2} Measurements for $\gamma$~Cru and $\mu$~Gem.}
\tablehead{
\colhead{$\lambda_\text{lab}$} &  \colhead{Multiplet} &  \colhead{$\tau_\text{rel}$\tablenotemark{a}} & \colhead{$RV_\text{em}$} & \colhead{$RV_\text{abs1}$} & \colhead{$RV_\text{abs2}$} & \colhead{$F_\text{surf}$}  
& \colhead{$RV_\text{em}$} & \colhead{$RV_\text{abs1}$} &  \colhead{$RV_\text{abs2}$} & \colhead{$F_\text{surf}$}  \\
\colhead{(\AA)} & \colhead{ } & \colhead{ } & \colhead{(km~s$^{-1}$)} & \colhead{(km~s$^{-1}$)} & \colhead{(km~s$^{-1}$)} & \colhead{(ergs cm$^{-2}$ s$^{-1}$)}  
& \colhead{(km~s$^{-1}$)} & \colhead{(km~s$^{-1}$)} & \colhead{(km~s$^{-1}$)} & \colhead{(ergs cm$^{-2}$ s$^{-1}$)} 
}
\startdata
\multicolumn{3}{c}{\ion{Fe}{2}} & \multicolumn{4}{c}{$\gamma$~Cru} & \multicolumn{4}{c}{$\mu$~Gem} \\
2331.307 &  35      &  614.4 &  $-$0.0 &  $-$13.5  &   \phantom{1}6.8 &  \phantom{1}523.8 &   $-$5.0 &  $-$11.0 & \phantom{1}9.6 &	303.2 \\
2332.800 &   3      & 2733.0 &  \phantom{$-$}0.6 &  $-$13.4  & \phantom{1}9.7 &  \phantom{1}590.4 &  $-$5.1 &  $-$12.8  &  \phantom{1}4.9  & 266.1 \\
2338.008 &   3      & 1562.7 &  $-$1.0 &  $-$13.3  & \phantom{1}8.4 &  \phantom{1}868.7 &   $-$6.3 & $-$11.0  &  \phantom{1}8.3 & 358.5 \\
2354.889 &  35      &  214.0 &  $-$1.7 &    \phantom{1}$-$8.9  & \nodata &  \phantom{1}352.2 &   $-$5.9 & \phantom{1}$-$9.7 & 10.5 & 141.9 \\
2362.020 &  35      &  477.7 &  $-$0.6 &    \phantom{1}$-$8.3  & \nodata &\phantom{1}553.4 & \nodata &\nodata & \nodata & \nodata \\
2364.825 &   3      & 1948.1 &  \phantom{$-$}1.4 &  $-$11.9  & 11.8 &  \phantom{1}355.9 & \nodata &\nodata & \nodata & \nodata \\
2366.593 &  35      &  273.9 &  \phantom{$-$}2.6 &    \phantom{1}$-$7.6 & \nodata  &  \phantom{1}326.4 & \nodata &\nodata & \nodata &\nodata \\
2485.076 &  34      &	 4.52 &  \phantom{$-$}2.9 & \nodata & \nodata &  \phantom{11}85.7 & \nodata &\nodata & \nodata & \nodata\\
2508.338 & \nodata  &	 0.0 & \phantom{$-$}1.4  & \nodata & \nodata &   \phantom{11}84.6 & \nodata &\nodata & \nodata & \nodata \\
2585.876 &   1      & 3952.0 &  \phantom{$-$}0.2 & $-$13.2 & 12.0 &  \phantom{1}753.5 & \nodata &\nodata & \nodata  & \nodata\\
2591.542 &  64      &  228.3 &  \phantom{$-$}2.6 & \phantom{1}$-$6.8 & \phantom{1}9.7 &  \phantom{1}299.1 &   $-$0.9 &  \phantom{1}$-$5.3 &  10.2 & 118.0 \\ 
2598.369 &   1      & 4425.1 &  \phantom{$-$}3.7 &  $-$10.7 & 13.1 &  \phantom{1}529.2 &   $-$1.0 &  $-$11.4 & \phantom{1}7.7 & 258.0\\
2599.395 &   1      &13534.0 & \phantom{$-$}3.3 &  $-$14.3 &11.4 &  1,038.9 &   $-$1.2 &  $-$11.5 &  \phantom{1}9.9 & 302.1 \\
2607.086 &   1      & 3537.0 &  \phantom{$-$}1.4 &  $-$11.9  & 10.9 &  \phantom{1}478.5 &   $-$2.3 &  $-$10.0 & 12.9 & 194.1 \\
2613.825 &   1      & 2018.8 &\nodata& \nodata &\nodata & \nodata &   $-$1.3 &  \phantom{1}$-$9.9 & \phantom{1}9.4 & \phantom{1}76.8 \\
2617.618 &   1      & 1380.3 &  $-$0.1 &  $-$10.8 & 12.0 &  \phantom{1}687.6   &   $-$3.8 &  \phantom{1}$-$8.1 &	11.7 & 230.2 \\
2619.075 & 171     &	 0.0 & \nodata & \nodata &\nodata & \nodata &   $-$1.5 &  \phantom{1}$-$0.3 & \nodata & \phantom{1}39.3 \\
2621.669 &   1      &  485.9 &   \phantom{$-$}3.9 & $-$10.7 & \phantom{1}8.7 &   \phantom{1}188.2 &  \phantom{$-$}1.5 &  \phantom{1}$-$9.7 & \phantom{1}7.3 & \phantom{1}84.2 \\
2625.664 &   1      & 1754.3 & $-$0.3 & $-$12.5 & \phantom{1}8.1 &  \phantom{1}915.4 &   $-$5.4 &  $-$10.5 & \phantom{1}9.6 &  361.4 \\
2628.291 &   1      & 1559.9 &  \phantom{$-$}6.1 & $-$11.9  & \phantom{1}8.4 &  \phantom{1}354.9 & \nodata &\nodata & \nodata & \nodata \\
2732.441 &  32      &	 1.5 &  \phantom{$-$}1.5 & \nodata & \nodata  & \phantom{1}390.0 & \nodata &\nodata & \nodata & \nodata \\
2736.968 &  63      &  194.1 &  $-$0.2 &   \phantom{1}$-$8.0 & \phantom{1}9.4 &  \phantom{1}824.5 &   $-$2.7 &  \phantom{1}$-$4.2 &\nodata  &  293.8 \\
2739.545 &  63      & 1738.1 & \phantom{$-$}2.0 & $-$11.7 & \phantom{1}8.9 & 1,779.6 &   $-$4.2 &  \phantom{1}$-$9.2 & \phantom{1}9.8 &  601.6 \\
2741.395 & 260      &     0.0 &  \phantom{$-$}1.9 & \nodata & \nodata &   \phantom{11}87.3 &  $-$2.2 &\nodata & \nodata & \phantom{1}25.4 \\
2755.733 &  62      & 2172.2 & \phantom{$-$}1.7 & $-$12.6 & \phantom{1}8.4 &  2770.2 & \nodata &\nodata & \nodata  & \nodata\\
2759.336 &  32      &	 1.3 &  \phantom{$-$}0.7 & \nodata & \nodata & \phantom{1}281.4 &  $-$2.4 & \nodata & \nodata & \phantom{1}70.3 \\
2761.813 &  63      &	83.4 &  \phantom{$-$}0.7 &    \phantom{1}$-$6.7 & \phantom{1}5.0  &  \phantom{1}765.8 &   $-$2.9 &  $-$10.7 &  \phantom{1}0.9 & 177.5\\
2772.719 &  63      &	21.6 &   \phantom{$-$}2.7 & \nodata & \nodata  &  \phantom{1}167.9 &   $-$2.6 &\nodata &	\nodata & \phantom{1}54.7 \\
2775.339 &  32      &	 0.6 &   \phantom{$-$}0.2 & \nodata & \nodata & \phantom{1}192.8 &   $-$2.6 & \nodata & \nodata & \phantom{1}45.5 \\
2861.168 &  61      &	 0.8 &\nodata& \nodata &\nodata & \nodata &  \phantom{$-$}0.6  &\nodata & \nodata &  \phantom{1}91.8 \\
2868.874 &  61      &	 6.2 &\nodata& \nodata &\nodata & \nodata &   \phantom{$-$}0.2 &\nodata & \nodata & 195.3 
\enddata
\tablenotetext{a}{The listed opacities are taken from \cite{carpenter95}.}
\end{deluxetable*}

The parametrized line modeling has an advantage over a simple multiple Gaussian fit in that it allows fitting optically thick lines where the wind absorption dominates the line profile. This is necessary to parameterize the stronger emission lines in which the central intensity approaches zero. The derived accuracy of the measured radial velocities is 0.5$-$3 km~s$^{-1}$, where the spread reflects the degree of line blending. 

As is the case for other non-coronal stars, the stronger self-reversed lines are better fit using two absorption components: one strong component with a significant blue-shift, and a second, weaker, red-shifted component. We have used the fits to measure the radial velocity of the emission and absorption components of the chromospheric lines. The uncertainty in a typical measurement of the absolute radial velocity of an individual line in a medium-resolution GHRS data frame is about 4~km~s$^{-1}$, including fitting uncertainties ($\sim$2~km~s$^{-1}$) and uncertainty in the absolute wavelength calibration of a single GHRS data frame ($\sim$3~km~s$^{-1}$). When more lines and/or more than a single GHRS spectrum are used, the uncertainties are reduced according to:

\begin{equation}\label{sigma}
\nonumber
\sigma_v=\sqrt{\left( \frac{3.0}{\sqrt{N_1}}\right)^2 + \left( \frac{2.0}{\sqrt{N_2}}\right)^2} \mathrm{km\,s^{-1}}
\end{equation}

where N$_1$ and N$_2$ are the number of used spectra and measured transition, respectively. Thus, if the lines are within a single data frame, and thus subject to the same uncertainty in zero-point of the wavelength scale, the minimum uncertainty, even with many lines is $\sim$3 km~s$^{-1}$. If the lines are spread over multiple data frames with different zero-point errors, the uncertainty is significantly reduced. The errors in the relative velocities of lines within a single spectrum are in the order of 0.5 km~s$^{-1}$.  We measured the surface flux, by fitting emission and absorption Gaussian profiles to the lines. Results are discussed in Section~\ref{results_empirical}.

\subsection{Semi-empirical modeling of the wind using the SEI formalism}\label{sei_form}
Semi-empirical models of chromospheres (for example in  $\alpha$ Tau) were developed by \citet{McMurry99}. To more precisely characterize the wind we model the UV line profiles using the 1D SEI code \cite{lam87} to solve the radiative transfer in a homogeneous, spherically expanding atmosphere using the Sobolev approximation, and explicitly including large-scale turbulence. 

The code requires the input of: the turbulent velocity in the envelope ($v_\text{turb}$) assumed constant with height, the wind velocity relation $v/v_\infty$ = (1-$R_\star/R)^{\beta}$ characterized by the acceleration parameter $\beta$, the optical depth of the modeled line as a function of the velocity, $\tau_{v}$, the opacity of the wind in each line, the collisional term (if any) in the source function, the underlying photospheric spectrum, and a lower boundary condition i.e. a chromospheric profile input to the base of the wind. 
 
We use the SEI approximation to compute line profiles for the wind absorptions seen over bright chromospheric emission lines and adjusted the input parameters to get the best fit to the observed line profiles. In addition to the very strong \ion{Mg}{2} resonance lines, we have a carefully-selected a set of \ion{Fe}{2} lines to sample a wide range of opacities and, hence, heights in the wind. The advantage of the SEI code is that it allows investigations of a broad range of parameters. It is difficult to simultaneously fit all of the lines presented with the same set of parameter values, and one converges to a relatively narrow range of wind parameters; $\beta$, $v_\text{turb}$, and $v_\infty$ are accurately derived in each case. When comparing the predicted lines with the observed line profiles, the physical parameters are derived from the line fitting. Our results are presented in Section~\ref{resultsSEI}. The SEI program does not solve the statistical equilibrium equations in the wind and, therefore, the optical-depth relation needs to be specified as input. For examples of applications of the SEI method on winds of O-stars and of planetary nebulae see e.g.~\citet{perinotto89, Groenewegen&Lamers89}.

Assumptions regarding wind temperatures, ionization ratios, and elemental abundances and, the mass-loss rates can be estimated from the inferred wind optical depths, adopting equation 29$b$ from \cite{olson82}:

\begin{equation}
\nonumber
\begin{split}
\dot{M} =& 8.70\times10^{-19}\mu\tau(v_\infty=0.5) \\
& \times \left(x^2\omega\frac{d\omega}{dx}\right) \frac{v_\infty^2 R_{\star}}{f \lambda_0 I A_{\text{E}}} \frac{U}{g e^{-q}}
\end{split}
\end{equation}												

where $x=R/R_{\star}$, $\omega=v(r)/v_\infty$, $q=E_l/kT$. While $I$ is the ionization fraction, $A_E$ the elemental abundance relative to hydrogen, $g$ the statistical weight, $f$ the oscillator strength, $E_l$ the lower energy level of the transition, $U$ the partition function, and $\lambda_0$ the central wavelength of the spectral line (in \AA); for further details see \citet{carpenter99a}.

\section{Results}\label{results}

\subsection{Wind parameters via empirical measurements} \label{results_empirical}

\begin{figure}[h]
\includegraphics[width=\columnwidth]{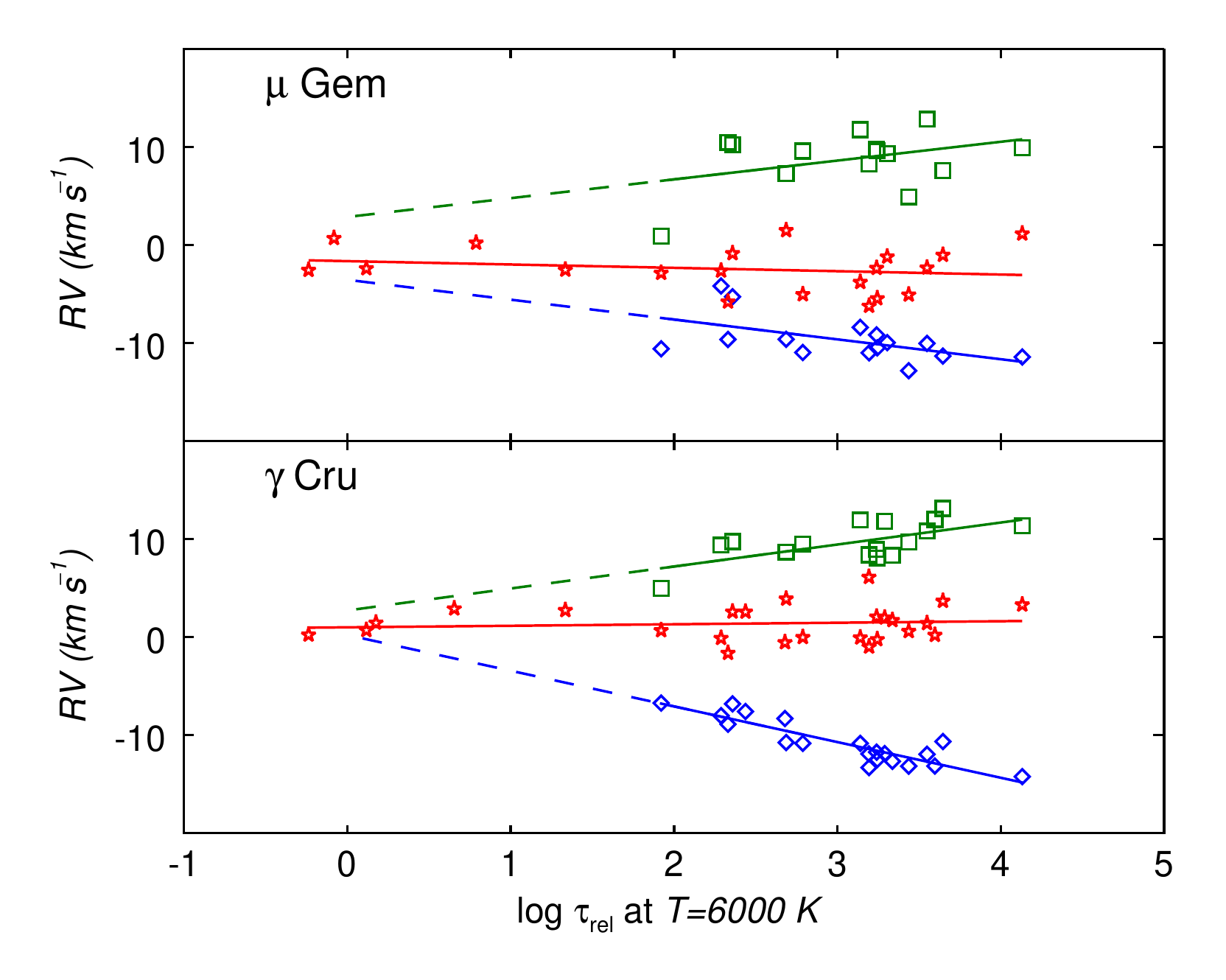}
\caption{The measured velocities of the chromospheric emission ($\ast$) and of the absorption components ($\diamond$ and $\square$) for the M3IIIb star $\mu$~Gem (upper) and the M3.4III $\gamma$~Cru (lower). The increase in the mean velocity of the wind absorption (as shown by the blue diamonds) with increasing line opacity reflects the acceleration of the wind with height. The chromospheric emission for $\gamma$ Cru is slightly red-shifted with a more rapidly accelerating wind than $\mu$~Gem.}
\label{object_accel} 
\end{figure}

Figure~\ref{object_accel} shows the measured velocities of the chromospheric emission and of the self-absorption as a function of line opacity for the two M-stars in this study. The opacity is used as a proxy for the atmospheric height. The measured first and second absorption components are listed in Table~\ref{FeII_MuGem_GamCru} (see also \citealp{carpenter95}). The monotonic increase (with relative optical depth) of the blue-shift of the dominant absorption component reflects the acceleration of the outflowing wind. The weaker, red-shifted, absorption component shows a redshift that increases with optical depth. The interpretation of this feature is ambiguous and depends on the geometry of the atmosphere. In the geometrically-thin (i.e.~plane-parallel) case, it represents an inflow of material; but if the formation region is very spherically extended, the feature could be caused by a simple monotonically increasing outflow, since the fluxes at these wavelengths would then be formed in regions preferentially {\it behind} the plane through the center of the star in the sky. Questions thence arise concerning the redward absorption components, as to the effects of occultation of photons by the stellar disk and whether the feature is really absorption or just a lack of emission. Detailed transfer calculations in spherical geometry are needed to answer these questions. Such transfer calculations are beyond the intent of this paper, and raise additional questions in themselves since they will be extremely sensitive to the adopted microturbulence and flow profiles. We intend to pursue such calculations in a later paper.  In this paper, we present our interpretation based on the assumption that the geometrical effects are minimal. In this case, the amount of the material involved in the downflow is substantially less than that in the upflow, as indicated by its much weaker absorption line strength. This would suggest that there may be a circulation pattern superposed on the dominant outflow, and that some of the material initially accelerated in the lower regions of the wind does not reach escape velocity and later returns toward the surface.

\begin{deluxetable*}{cccccccc}
\tablecolumns{8}
\tabletypesize{\scriptsize}
\tablecaption{Stellar Parameters and Empirical Wind Measures.\label{starsparams}}
\tablewidth{0pc}
\tablehead{
\colhead{Star} & \colhead{Type} & \colhead{$T$($T_\sun$)} & \colhead{$L$($L_\sun$)} & \colhead{$\log g$} & \colhead{$R$ ($R_\sun$)} & \colhead{Max $V_\text{abs}$\tablenotemark{a} (km~s$^{-1}$)} & \colhead{Mean $V_\text{abs}$\tablenotemark{b} (km~s$^{-1}$)}
}
\startdata
\multicolumn{8}{l}{K-stars (prev. works)} \\ 
$\alpha$~Tau  & K5 III  & 3898$\pm$30\tablenotemark{c}  & 394$\pm$15\tablenotemark{c} & 1.25 &  44 & 40 & 1$-$25 \\
$\gamma$~Dra  & K5 III   & 3985$\pm$45\tablenotemark{c} & $535\pm25$\tablenotemark{c} & 1.50 &  49 & 80 & 5$-$50\tablenotemark{d} \\
                           & hybrid & \nodata   & \nodata & \nodata & \nodata & \nodata & 50$-$80\tablenotemark{e} \\
$\lambda$~Vel   & K5 Ib    & 3820\tablenotemark{f} &  7943\tablenotemark{f}  & 0.64 & 210 & 60 & 10$-$25 \\ \hline
\multicolumn{8}{l}{M-stars (this work)}  \\ 
$\mu$~Gem     & M3 IIIab & 3675\tablenotemark{h}  &  2754\tablenotemark{i}   & 1.5\tablenotemark{j}  & 230.4\tablenotemark{k} & 25 & 9$-$13  \\
$\gamma$~Cru  & M3.4 III & 3689\tablenotemark{l} & 758 & 2.00 & 120 & 25 & 6$-$14\tablenotemark{m}   \\ 
              & \nodata & \nodata  & \nodata &  \nodata &\nodata & 50 & 6$-$40\tablenotemark{n} \\
\enddata 
\tablenotetext{a}{The Max. V$_\text{abs}$ is a measure of the V$_\infty$+V$_\text{turb}$ in the wind if lines of sufficient opacity.}
\tablenotetext{b}{The Mean V$_\text{abs}$ is the centroid of the wind absorption features $-$ the range reflects that observed for lines of different opacity.}
\tablenotemark{c}{\cite{robinson98a}}
\tablenotetext{d}{Primary (low-velocity) wind component.}
\tablenotetext{e}{Secondary (high-velocity) wind component.}
\tablenotemark{f}{\cite{carpenter99a}}
\tablenotemark{h}{\cite{wood16}}
\tablenotemark{i}{\cite{mallik99}} 
\tablenotemark{j}{\cite{massarotti08}} 
\tablenotemark{k}{Calculated with the Stefan-Boltzmann law.} 
\tablenotemark{l}{\cite{mcdonald17}}
\tablenotemark{m}{During the majority of observational epochs (normal wind).}
\tablenotemark{n}{During strong/high-opacity wind epoch (April 1978).}
\end{deluxetable*}

As for the previously studied K-stars the mean velocity of the emission is approximately at rest with respect to the photosphere, while the mean outflow velocity of the wind absorptions increases with opacity. This indicates that the line photons are created in a region at rest with respect to the star, below the region of the wind acceleration.  


We do not see the acceleration in its early stages because the wind absorptions in the weak lines, which could sample those low altitudes (velocities), have insufficient total opacity. Table~\ref{FeII_MuGem_GamCru} shows the results of those fits.

Fitting emission and two absorption gaussian profiles to the lines, and calculating velocities for the center of the gaussian profiles, we derived net integrated surface fluxes values by subtracting the two absorption components from the emission integrated flux. We calculated extinction factors following \cite{cardelli89}, and using values of $A_\text{v}$ and $R_\text{v}$ from \cite{gontcharov18}. For $\mu$~Gem those are: $A_\text{v}$= 0.22, $R_\text{v}$=2.99; while for $\gamma$~Cru: $A_\text{v}$=0.16, $R_\text{v}$=3.09. To calculate the surface fluxes we adopt angular diameter values of: 24.7~mas for $\gamma$~Cru \citep{glindemann01}, while for $\mu$~Gem we used an average of the limb-darkened values from the CHARM2 catalogue: $\theta_\text{LD}$=13.45~mas \citep{richichi05}. Results of the surface fluxes calculations are presented in Table~\ref{FeII_MuGem_GamCru} for the two investigated stars. 

Table~\ref{starsparams} lists the basic parameters of the program and compare stars along with two empirical measurements of the wind speed, the maximum velocity from line center at which wind absorption is seen in any line in the spectrum, and the range of mean velocities seen in lines of different strength throughout the spectrum. The radius of $\mu$~Gem is given by applying the Stefan-Boltzmann law ($L=4\pi R^2 \sigma T^4$) to the ratio of luminosities of $\mu$~Gem/$\gamma$~Cru, consequently $R_{\mu\text{Gem}}$=230.4~R$_\sun$, and $R_{\gamma\text{Cru}}$=120.0~R$_\sun$.

\subsection{SEI fits to {\it HST}/GHRS spectra}\label{resultsSEI}
\begin{figure}[!h]
\includegraphics[width=\columnwidth, bb=262 3 769 761]{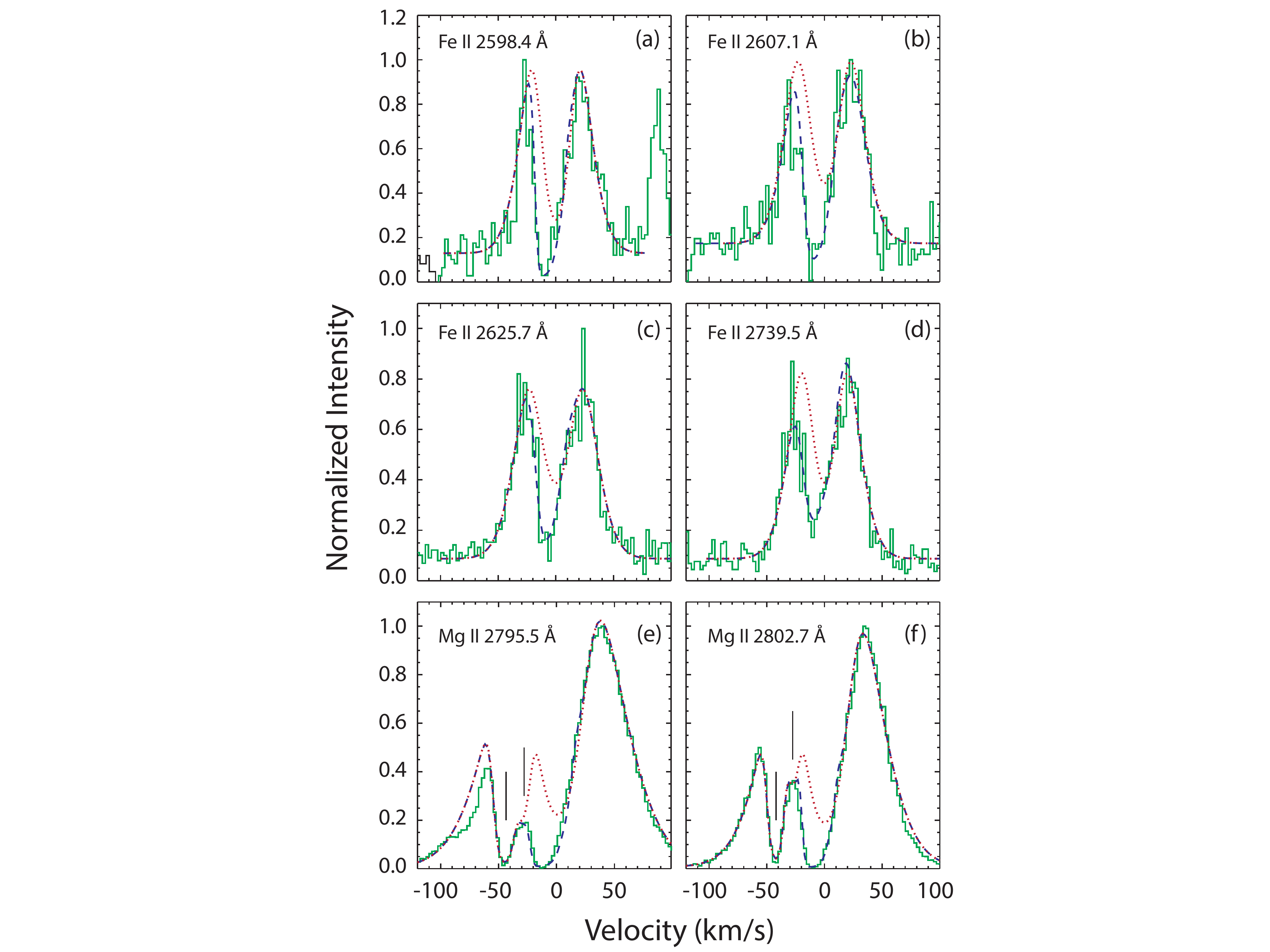}
\caption{Sample of SEI fits to lines representing a wide range of optical depths in the spectra of $\mu$~Gem. The full green line shows the GHRS observed spectrum; the red dotted line is the chromospheric profile input to the base of the wind; the blu dashed line is the calculated profile. The two dark vertical lines over the Mg~II profiles at $2795.5~$\AA (panel e) and $2802.7~$\AA (panel f) underline two interstellar features (see Table~\ref{seitable}). \label{mg_egseifits}}
\end{figure}

\begin{figure}[!h]
\centering
\includegraphics[width=\columnwidth, bb=262 3 763 759]{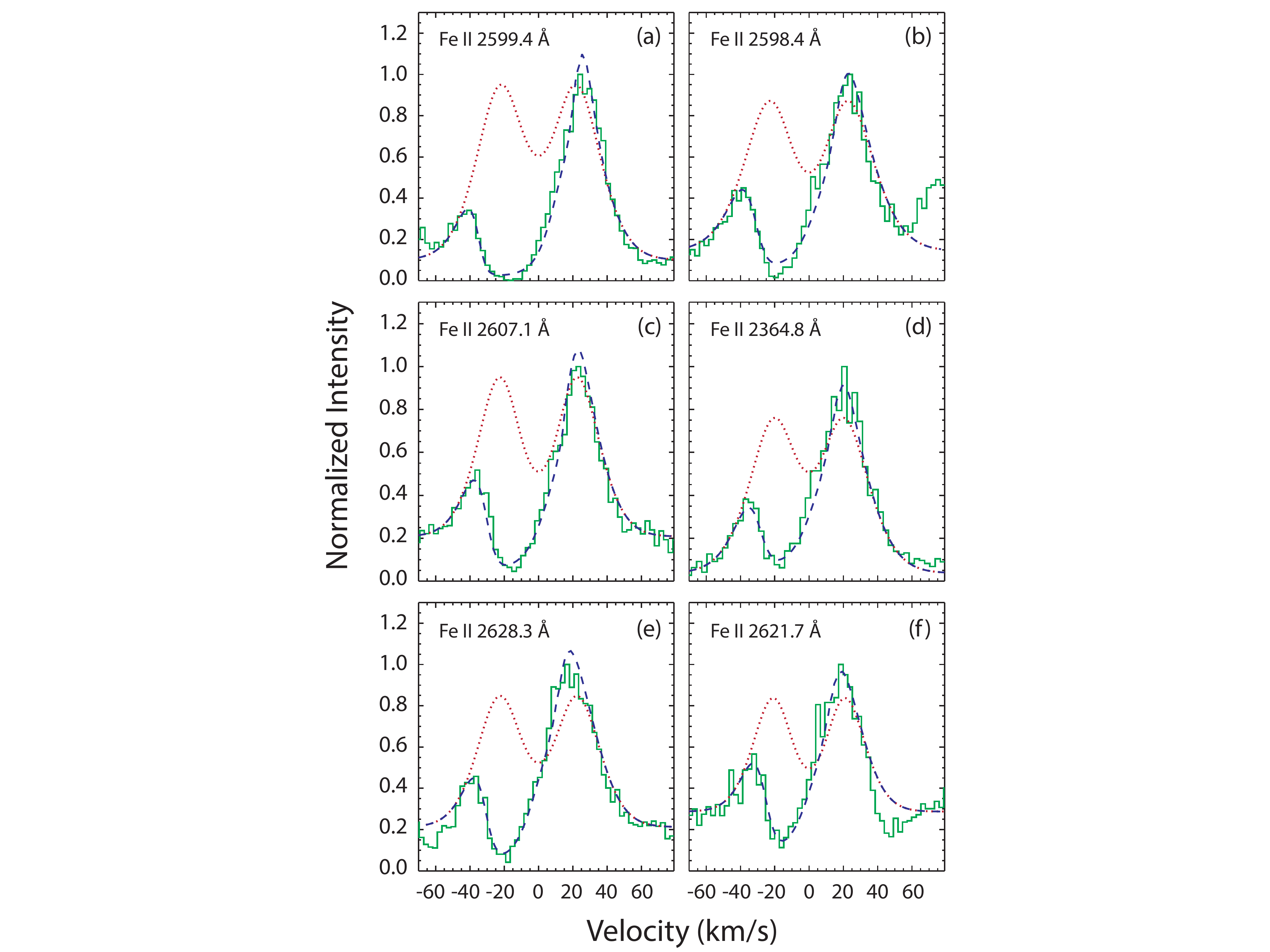}
\caption{Sample of SEI fits to lines representing a wide range of optical depths in the spectra of $\gamma$~Cru. The full green line shows the GHRS observed spectrum; the red dotted line is the chromospheric profile input to the base of the wind; and the blu dashed line is the calculated profile. \label{gc_egseifits}}
\end{figure}

The method of the SEI fit described in Section~\ref{sei_form} was applied to the {\it HST}/GHRS data of $\mu$~Gem and $\gamma$~Cru. Figures~\ref{mg_egseifits} and~\ref{gc_egseifits} show a sample of SEI fits to emission lines in $\mu$~Gem and $\gamma$~Cru, respectively.

\begin{figure}
\includegraphics[width=\hsize, bb=77 154 693 504, angle=180]{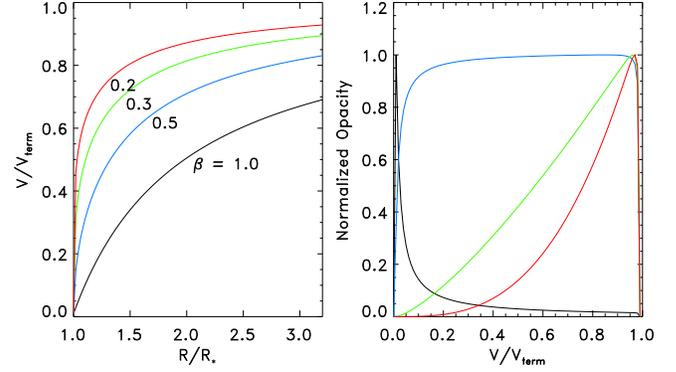}
\caption{The dependence of the velocity vs. radius and normalized opacity vs. velocity relations on the $\beta$ wind parameter.  \label{betacurves} }
\end{figure}

The dependence of the wind velocity vs.~height and the opacity vs. velocity for various values of the wind acceleration parameter $\beta$ are shown in Figure~\ref{betacurves}. The parameters of the fit to each emission line are given in Table~\ref{seitable}, for $\mu$~Gem and $\gamma$~Cru.

From the fits, we are able to estimate the wind parameters, listed in Table~\ref{sei-results}, inferred from our SEI modeling of the complete set of lines, for both the program and comparison stars (see modeling description in Section~\ref{sei_form}).

\begin{deluxetable*}{cc | cccc | cccc}
\tablecolumns{10}
\tablewidth{0pc}
\tablecaption{Results of SEI fits to GHRS data of $\gamma$ Cru and $\mu$ Gem. \label{seitable} }
\tablehead{
\colhead{$\lambda_\text{lab}$ (\AA)} & \colhead{Multiplet} & \colhead{$\beta$} & \colhead{$v_\text{turb}$/$v_{\infty}$} & \colhead{$v_{\infty}$ (km~s$^{-1}$)} & \colhead{$\log\tau_\text{wind}$} 
&  \colhead{$\beta$} & \colhead{$v_\text{turb}$/$v_{\infty}$} & \colhead{$v_{\infty}$(km~s$^{-1}$)} & \colhead{$\log\tau_\text{wind}$}}                                                                            
\startdata 
\multicolumn{2}{c}{\ion{Fe}{2}} & \multicolumn{4}{c}{$\gamma$ Cru} & \multicolumn{4}{c}{$\mu$ Gem} \\
2585.867 & 1 & 0.7 & 0.65 & 18 & 25 & \nodata & \nodata & \nodata & \nodata  \\
2598.369 & 1 & 0.7 & 0.7 & 18 & 30 & 0.6 & 0.8 & 11 & 14 \\
2599.395 & 1 & 0.8 & 0.7 & 19 & 50 & 0.4 & 0.7 & 14 & 20  \\
2607.086 & 1 & 0.7 & 0.7 & 18 & 12 & 0.6 & 0.8 & 11 & 7 \\
2617.618 & 1 & 0.7 & 0.7 & 17 & 9 & 0.6 & 0.8 & 11 & 4.5 \\
2621.669 & 1 & 0.6 & 0.6 & 19 & 6 &  \nodata & \nodata & \nodata & \nodata  \\
2625.664 & 1 & 0.7 & 0.6 & 17.5 & 8.5 &  0.6 & 0.8 & 11 & 4.0  \\
2628.291 & 1 & 0.7 & 0.45 & 23 & 8 &  0.6 & 0.8 & 14 & 4  \\ 
2332.800 & 3 & 0.8 & 0.7 & 17 & 22 &  0.6 & 0.8 & 11 & 10 \\
2338.008 & 3 & 0.7 & 0.7 & 17 & 10 &  0.6 & 0.8 & 11 & 5    \\
2364.829 & 3 & 0.7 & 0.7 & 19 & 9 &  \nodata & \nodata & \nodata & \nodata\\ 
2331.301 & 35 & 0.7 & 0.7 & 17 & 14 & 0.6 & 0.8 & 9 & 8  \\
2354.889 & 35 & 0.7 & 0.7 & 19 & 3.5 &  0.6 & 0.8 & 11 & 2.5   \\
2362.020 & 35 & 0.7 & 0.7 & 19 & 4  &  \nodata & \nodata & \nodata & \nodata\\
2366.593 & 35 & 0.7 & 0.7 & 19 & 1.5 &  \nodata & \nodata & \nodata & \nodata\\ 
2755.733 & 62 & 0.7 & 0.7 & 20 & 7 & 0.6 & 0.8 & 13 & 3  \\ 
2736.968 & 63 & 0.7 & 0.7 & 19 & 2 &  0.6 & 0.8 & 8 & 1  \\
2739.545 & 63 & 0.7 & 0.7 & 19 & 7.5 &  0.6 & 0.8 & 14 & 3  \\ 
2591.542 & 64 & 0.7 & 0.7 & 19 & 2.5 &  0.6 & 0.8 & 10 & 2.0 \\
2593.722 & 64 & 0.7 & 0.6 & 19 & 5.5 &  \nodata & \nodata & \nodata & \nodata \\ \hline 
\multicolumn{2}{c}{\ion{Mg}{2}} & \multicolumn{4}{c}{$\gamma$ Cru} & \multicolumn{4}{c}{$\mu$ Gem\tablenotemark{a}}\\
2795.523 & 1 & 0.8 & 0.7 & 21.5 & 80 &  0.6 & 0.6 & 14 & 30 \\
2802.698 & 1 & 0.8 & 0.7 & 19 & 60 &  0.6 & 0.6 & 13 & 30 \\
\enddata 
\tablenotetext{a}{Two interstellar absorption features are seen in the Mg II h and k lines (see also e.g.~\cite{carpenter97b, malmut97}: \\
1. RV=12.2 (k), 11.7 (h) km s$^{-1}$, equal to 42.26 (k), and 42.76 (h) relative to the stellar radial velocity (see Fig.~\ref{mg_egseifits}); FWHM=0.13 (k), 0.11 (h) \AA \\
2. RV=27.2 (k), 26.0 (h) km s$^{-1}$ equal to 27.26 (k), and 28.46 (h) relative to the stellar radial velocity (see Fig.~\ref{mg_egseifits}); FWHM=0.12 (k), 0.11 (h) \AA \\}
\end {deluxetable*}

\begin{deluxetable}{lcccc}
\tablecolumns{5}
\tablewidth{0pc}
\tablecaption{Results of SEI Modeling. \label{sei-results}}
\tablehead{
\colhead{Star} & \colhead{$\beta$} & \colhead{$v_\infty$} & \colhead{$v_\text{turb}$} & \colhead{Mass-Loss Rate} \\
\colhead{} & \colhead{} & \colhead{(km~s$^{-1}$)} & \colhead{(km~s$^{-1}$)} & \colhead{($\times$10$^{-11}M_{\sun}$/yr)}}
\startdata
\multicolumn{5}{l}{M$-$stars (this paper)} \\ 
$\mu$~Gem     & 0.6  & 11 &  9   & 7.4    \\
$\gamma$~Cru  & 0.7  & 19 & 14   & 45     \\
\hline
\multicolumn{5}{l}{K$-$stars\tablenotemark{a}}  \\ 
$\alpha$~Tau  & 0.6  & 30 & 24   & 1.4      \\
$\gamma$~Dra  & 0.6  & 30 & 24   & 0.14\tablenotemark{b} \\
              & 0.35 & 67 & 12   & 1.20\tablenotemark{c} \\
$\lambda$~Vel & 0.9  & 31 & 9-21 & 300    
\enddata
\tablenotetext{a}{$\alpha$~Tau and $\gamma$~Dra from \cite{robinson98a}; $\lambda$~Vel from \cite{carpenter99a}.}
\tablenotetext{b}{weak, secondary wind component}
\tablenotetext{c}{strong, primary wind component}
\end{deluxetable}

SEI models of the outflowing winds indicate that $\mu$~Gem has, in general, a weaker wind, in terms of turbulent and terminal velocity, than $\gamma$~Cru.  This is consistent with expectations, given the higher surface gravity and lower luminosity class of $\mu$~Gem (see \citealp{rau18a}). 

Our results show that for $\mu$~Gem the wind opacity in each self-reversed emission line is significantly smaller than in $\gamma$~Cru. Also, the turbulent velocity in the wind are smaller in  $\mu$~Gem (9 km~s$^{-1}$) vs.~$\gamma$~Cru (14 km~s$^{-1}$). The same is true for the terminal velocity: 11~km~s$^{-1}$\ in $\mu$~Gem vs.~19~km~s$^{-1}$\  in $\gamma$~Cru; and for the corresponding mass-loss rate: 7$\times$10$^{-11}M_{\sun}$/yr for $\mu$~Gem vs.~45$\times$10$^{-11}M_{\sun}$/yr for $\gamma$~Cru (see Table~\ref{sei-results}).

Table~\ref{cf-results} compares the mass-loss rate calculations resulting from the present work, and from other techniques, with the findings of earlier spectral-type K giants $\alpha$~Tau, $\gamma$~Dra, and $\lambda$~Vel. This comparison shows that the winds from K giants are much faster and terminal wind velocities of K giants are greater by a factor of two. However, the rate of wind acceleration is comparable ($\beta$ = 0.6 vs.~0.7) in the two stars. The measurements suggest that M giants have slower but more massive winds.

\begin{deluxetable}{lrrrrr}
\tablecolumns{6}
\tablewidth{0pc}
\tablecaption{Comparison with other Mass-Loss Estimates ($\times$10$^{-11}M_\sun$/yr). \label{cf-results}}
\tablehead{
\colhead{Star}  &  \colhead{SEI\tablenotemark{a}} & \colhead{Optical\tablenotemark{b}} &  \colhead{Radio\tablenotemark{c}} & \colhead{J\&S\tablenotemark{d}} & \colhead{K\&R\tablenotemark{e}}}
\startdata
\multicolumn{6}{l}{M$-$stars} \\
$\mu$~Gem     & 7.4  &     \nodata     &    \nodata     &   \nodata  &   0.20   \\
$\gamma$~Cru  & 45  &  \nodata & \nodata &  7  & 0.03 \\ \hline
\multicolumn{6}{l}{K$-$stars} \\
$\alpha$~Tau  & 1.4  & \nodata & 6.5  & 80  & 0.03 \\
$\gamma$~Dra  & 1.2  & \nodata & \nodata & 35  & 0.02 \\
$\lambda$~Vel & 300 & $<$800 & \nodata  & 600 & 1.26 
\enddata
\tablenotetext{a}{See Table~\ref{sei-results}}
\tablenotetext{b}{\cite{hag83} using optical data}
\tablenotetext{c}{\cite{drake86}, as quoted by J\&S}
\tablenotetext{d}{\cite{jud91} using an empirical AGB/RGB relation}
\tablenotetext{e}{\cite{kud78} using an empirical all stars relation}
\end{deluxetable}

\section{Discussion} \label{discussion}
\subsection{Winds and stellar parameters}

Several observational studies have demonstrated the important role of Alfv\'en waves propagating in the lower solar atmosphere to deliver the energy to the corona, as observed by \citet{jess09}.   

\cite{verdini10, vf12, suzuki07, suzuki13} demonstrated the importance of propagation and dissipation of waves in the solar coronal heating through turbulent cascade of Alfv\'en waves; however, this mechanism is not at work in cool, evolved stars, as the surface gravity of the Sun is much higher than for M-giant stars. These papers indeed predominantly describe the wind acceleration features due to turbulent heating caused by counter-propagating  Alfv\'en waves in solar coronal loops, and do not address the wave dissipation and chromospheric heating. \cite{airapetian00, airapetian10, airapetian15} have shown that winds from late-type giant stars can be successfully modeled via Alfv\'en waves driven acceleration, due to the propagation of waves in partially ionized atmosphere, and their reflection in the gravitationally stratified atmospheres. 

Therefore, to understand the difference in wind dynamics in $\mu$~Gem and $\gamma$~Cru, we need to examine the atmospheric properties of these two stars.  As shown in Table~\ref{starsparams},  effective temperatures $T_{\text{eff}}$ of both observed stars are similar, while $\mu$~Gem is 3$-$4 times more luminous than $\gamma$~Cru, suggesting that its surface area is correspondingly larger. Thus, if both stars generate winds at the same efficiency then the mass-loss of $\mu$~Gem should be expected to be 3$-$4 greater than for $\gamma$~Cru. However, Table~\ref{sei-results} indicates that the wind of $\mu$~Gem is a factor $6$ weaker. This suggests that, assuming a spherically symmetric wind, the wind mass-loss rate per unit surface area of $\mu$~Gem is a factor of 20 smaller than for $\gamma$~Cru.

The chromospheric thickness of $\mu$~Gem, determined from its pressure scale height $H$, given by $H\propto T_e/g$, is by a factor of 3 larger than in $\gamma$~Cru. Thus, the characteristic frequency of acoustic waves generated at the atmosphere, which scales as $v_s/H$, is correspondingly 3 times lower. 

A possible mechanism of Alfv\'en waves excitation at the wind base is acoustic shocks that dissipate and heat the stellar chromosphere \citep[see e.g.][]{judge93}. As the Alfv\'en waves propagate upward in the atmosphere, $\mu$~Gem experiences reflections at much lower heights in the atmosphere, and drive slower winds at lower speeds, compared to $\gamma$~Cru. Since the amplitudes of upward propagating waves increase as $\rho^{-\frac{1}{4}}$, they are expected to be correspondingly smaller, as lower frequency waves get reflected from lower heights. This is consistent with the measurements of non-thermal broadening of \ion{Fe}{2} intercombination chromospheric lines (see Table~\ref{sei-results}). 

\begin{figure}
\centering
\includegraphics[width=\hsize, bb=0 0 612 721, angle=0]{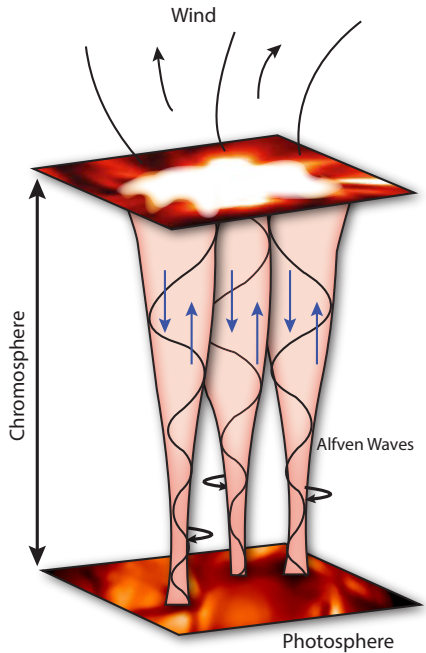}
\caption{Schematic representation of the Alfv\'en wave driven winds in late-type stars. Alfv\'en waves are generated in cool, evolved stars by direct perturbations of magnetic fields at the wind base due to transverse velocity perturbations, driven by nonlinear sound waves propagating upward from the stellar photosphere into the chromosphere. \label{alfen}}
\end{figure}

While detailed MHD modeling is required to reproduce the mass-loss rates, turbulent, and terminal wind velocities implied from observations (Rau, Airapetian, et al., in prep.), our {\it HST}/GHRS observations can be interpreted within the framework of the Alfv\'en wave driven winds, as shown in Figure~\ref{alfen} (see  \citealp{airapetian10}). Alfv\'en waves that drive stellar winds are generated at the top of the chromosphere by acoustic shocks. The latter are formed due to photospheric convection and/or pulsation mechanisms in evolved K or M-giant stars. Alfv\'en waves are efficiently dissipated well below the top of the stellar chromosphere where the wind acceleration is initiated. From our measurements of the surface fluxes, the chromosphere is stronger in $\gamma$~Cru than in $\mu$~Gem.

\subsection{Robustness of SEI modeling}\label{discussSEI}
Examples of our line fits in the {\it HST}/GHRS spectra of $\mu$~Gem and $\gamma$~Cru are shown in Figures~\ref{mg_egseifits} and \ref{gc_egseifits} and discussed in Section~\ref{resultsSEI}.  The figures show that we are able to obtain good fits to lines with different opacity with a self-consistent set of wind parameters. The fits constrain $\beta$ to $\pm$0.2, the $v_\infty$ to  $\pm$2 km~s$^{-1}$\ , the turbulence to $\pm$5 km~s$^{-1}$, and the mass-loss rates to about $\pm$30~\%.  However, the line opacity is difficult to constrain, in particular,   for high opacity, often saturated, lines.

The validity of the SEI modeling to cool giant and supergiants has been assessed by comparison with co-moving frame CRD calculation and is discussed in \citet{carpenter99a}. Their comparison shows that the SEI profiles are a good approximation over the entire line profile up to $\beta\approx$~1. At a higher value of $\beta$ the SEI profiles are still sufficiently accurate compared to observed profiles outside the red wing of the line cores.  

The advantage of the SEI code is that it is computationally fast and allows a much more efficient initial exploration of wind parameters space. The primary assumptions/limitations of the SEI technique are: a uniform wind temperature; treating the wind as a pure-scattering medium (and thus assuming that all the emerging photons are created in the chromosphere below the initiation of the wind flow); and a two-level approximation for the transitions. The impact of the latter appears minimal when fitting normalized profiles, especially in the wind-absorption region as shown in \cite{carpenter99a}. We assume a $\beta$ power velocity law, no photon creation in wind, and spherical symmetry. 

The uniform wind temperature may preclude application to M-supergiants and to the coolest giants (see e.g.~\citealp{carpenter99a}). Indeed, the uniform wind temperature and the treatment of the wind as a pure-scattering medium are true for the K-stars and the M-giants considered here. But we can not extend this to higher luminosities in the M-supergiants, because their winds appear to begin their acceleration well-within the region of photon creation. Any attempt to push the application of the SEI technique too far becomes quickly evident, as it is not possible to fit the full variety of \ion{Fe}{2} and \ion{Mg}{2} lines seen in the 2200$-$3200 \AA\ spectrum with a consistent set of wind parameters. 

The two analysis techniques presented in this paper are complementary, and we do not use them to derive and estimate the same parameters. The empirical measurements do not depend on a model, but are direct measures. The SEI modeling on the other hand, is needed to estimate the mass-loss rate (which we can not measure directly) and to derive a more detailed estimate of the detailed shape of the wind flow vs.~height/opacity. Indeed, the straight line measurements, as done in the empirical modeling, average over a range of opacities/heights for each line/data point and does not permit that.

\subsection{Comparison with previous works}\label{previousworks}
For the two stars studied in the present work, $\mu$~Gem and $\gamma$~Cru, a comparison of computed and observed UV emission line profiles containing overlying wind absorption indicate that the line photons are created in a region approximately at rest with respect to the photosphere. The winds are rather rapidly accelerating ($\beta\la$1), with turbulent velocities of 10$-$20 km~s$^{-1}$, and terminal velocities of 20$-$30 km~s$^{-1}$\ (for the non-coronal stars) and 67 km~s$^{-1}$\ (for the hybrid star). The mass-loss rates are of 10$^{-11}M_{\sun}$/yr for the K giants $\alpha$~Tau and $\gamma$~Dra, 4$\times$10$^{-10}M_{\sun}$/yr for the M giants $\mu$~Gem and $\gamma$~Cru, and 3$\times$10$^{-9}M_{\sun}$/yr for the K supergiant $\lambda$~Vel. Table~\ref{cf-results} shows that $\lambda$~Vel mass-loss rate is consistent with previous optical upper limit, but disagrees with current radio model for all stars, the derived mass-loss rates are within about an order of magnitude of the \cite{jud91} semi-empirical relation, but widely discordant with the \cite{kud78} relation.
 



\section{Conclusions}\label{conclusion}
In this paper we studied the two M giant stars $\mu$~Gem and $\gamma$~Cru, modeling their chromospheric contribution with two approaches: empirical modeling, and SEI modeling. We derived their wind parameters, and note that $\mu$~Gem has a higher mass and luminosity, a lower surface gravity, and a weaker wind and chromosphere than $\gamma$~Cru, suggesting that $\mu$~Gem is the more evolved star. 

If we consider the surface flux as a measurement of the strength of the chromosphere, we can conclude that the chromosphere of $\gamma$~Cru appears stronger in comparison to $\mu$~Gem.

We compared the results of the present work with previous ones on different type (K giant and supergiant) stars (see Section~\ref{discussion}). We observe that for the two M giants in this study, the terminal velocity of the wind is slower than in the earlier giant stars, but the mass-loss rate is considerably higher.

To understand the full dynamics of those winds, simulations of the chromospheric heating mechanism are necessary. We are thus planning to implement a more sophisticated modeling (Rau, Airapetian, et al., in prep.), using parameters derived with the SEI modeling as starting point for magnetohydrodynamic calculations. 

\bibliographystyle{apj}
\bibliography{master_papermugamma}

\acknowledgments This research is sponsored in part by NASA through STScI grants for {\it HST}\ GO program 5307. The authors would like to thank Gladys V.~Kober for the help on some calculations.
\software{CALHRS, SEI (Lamers et al. 1987)}
\end{document}